\documentstyle[emulateapj,psfig]{article}
\begin{document}

\title{Testing the Universal Structured Jet Models of Gamma-Ray Bursts by BATSE Observations}

\author{Xiaohong Cui$^{1,4}$,  Enwei Liang$^{1,2,3}$, and Ruijing Lu$^{1,2,4}$}
\affil{$^1$National Astronomical Observatories/Yunnan Observatory,
Chinese Academy of Sciences, Kunming 650011, China\\
$^2$Physics Department, Guangxi University, Nanning 530004, China\\
$^3$Physics Department, University of Nevada, Las Vegas, NV89154, USA; Email: lew@physics.unlv.edu\\
$^4$The Graduate School of the Chinese Academy of Sciences, Beijing, China }

\begin{abstract}
Assuming that the observed gamma-ray burst (GRB) rate as a function of redshift is proportional to a corrected star
formation rate, we derive the empirical distribution of the viewing angles of long BATSE GRBs, $P^{\rm em}(\theta)$,
and the distribution of these bursts in the plane of $\theta$ against redshift, $P^{\rm em}(\theta, \ z)$, by using a
tight correlation between collimation-corrected gamma-ray energy ($E_{\gamma}$) and the peak energy of $\nu F_{\nu}$
spectrum measured in the rest frame ($E_{\rm p}^{'}$). Our results show that $P^{\rm em}(\theta)$ is well fitted by a
log-normal distribution centering at $\log \theta/{\rm rad}=-0.76$ with a width of $\sigma_{\log \theta}=0.57$. We test
different universal structured jet models by comparing model predictions, $P^{\rm th}(\theta)$ and $P^{\rm th}(\theta,
\ z)$, with our empirical results. To make the comparisons reasonable, an ``effective" threshold, which corresponds to
the sample selection criteria of the long GRB sample, is used. We find that (1) $P^{\rm th}(\theta)$ predicted by a
power-law jet model is well consistent with $P^{\rm em}(\theta)$, but $P^{\rm th}(\theta, \ z)$ predicted by this model
is significantly different from $P^{\rm em}(\theta, \ z)$; (2) $P^{\rm th}(\theta, \ z)$ predicted by a single Gaussian
jet model is more consistent with $P^{\rm em}(\theta, \ z)$ than that predicted by the power-law jet model, but $P^{\rm
th}(\theta)$ predicted by this model rapidly drops at $\theta>0.3$ rad, which greatly deviates $P^{\rm em}(\theta)$;
and (3) both $P^{\rm th}(\theta)$ and $P^{\rm th}(\theta, \ z)$ predicted by a two-Gaussian jet model are roughly
consistent with our empirical results. A brief discussion shows that cosmological effect on the $E_{\gamma}-E_{\rm
p}^{'}$ relation does not significantly affect our results, but sample selection effects on this relationship might
significantly influence our results.

\end{abstract}

\section{Introduction}               % Introduction goes below.
The phenomenon of gamma-ray burst (GRB) is still of a great mystery, although significant progress has been made in the
recent decade (see reviews by Fishman \& Meegan 1995; Piran 1999; van Paradijs, Kouveliotou, \& Wijers 2000; Cheng \&
Lu 2001; M\'{e}sz\'{a}ros 2002; Zhang \& M\'{e}sz\'{a}ros 2004; Piran 2005). It is widely believed today that the
central engines of GRBs power conical ejecta (jet) to produce the observed GRBs and their afterglows. Sharp breaks
and/or quick decays of afterglow light curves are regarded as evidence of jets (e.g., Rhoads 1999; Sari, Piran, \&
Halpern 1999). Such evidence is rapidly growing up in the recent years (e.g., Bloom, Frail, \& Kulkarni 2003 and the
references therein).

The structure of GRB jet is currently under heavy debate. Uniform jet model (e.g., Rhoads 1999; Frail et al. 2001) and
universal structured jet (USJ) model (e.g., M\'{e}sz\'{a}ros, Rees \& Wijers 1998; Dai \& Gou 2001; Rossi, Lazzati, \&
Rees 2002; Zhang \& M\'{e}sz\'{a}ros 2002) are two currently competing models. In the framework of the uniform jet
model, jet opening angle is assumed to be different from burst to burst, but energy distribution within a jet is
uniform. In the scenario of the structured jet model, it is assumed that the energy (and/or bulk Lorentz factor)
distribution within a jet is a function of the angle measured from jet axis. The USJ models suggest that all GRB jets
have the same geometric structure and the same energy distribution within jets. In this model, the observational
diversity of GRB population is resulted only from different viewing angles. Since there are significant dispersions in
GRB data, quasi-universal structured jet models are also proposed (Zhang et al. 2004a; Lloyd-Ronning, Dai, \& Zhang
2004; and Dai \& Zhang 2005). Numerical simulations of jet propagation within collapsar context by Zhang, Woosley, \&
Heger (2004b) showed that the jet structure seems to be quite complicated, especially when the part of $\theta>0.1$ rad
is taken into account. Such a jet structure is difficult to model by a single component. A jet structure with two or
more components is  likely to be more real. The observations of GRB 030329 seem to favor such a jet model (Berger et
al. 2003). The afterglow lights of this burst show two temporal breaks---one occurs at 0.5 day after the burst trigger
in optical afterglow light curve (Price et al. 2003) and the other happens at 9.5 days in the radio afterglow light
curve (Berger et al. 2003). Berger et al. (2003) argued that the interpretation for the two breaks requires a
two-component jet with $\theta_{c,1}=0.1$ rad and $\theta_{c,2}=0.3$ rad. Millimeter observations of this burst also
support this jet model (Sheth et al. 2003). Liang \& Dai (2004) explained the plausible bimodal distribution of the
observed peak energy of $\nu F_{\nu}$ spectra ($E_{\rm p}$) with the similar jet structure model. Huang et al. (2004)
and Peng, K\"{o}nigl, \& Granot (2004) investigated optical afterglows from such a jet model.

Afterglow observations present more detailed information, but in the late afterglow phase jet structure signatures may
be washed out. Combination of prompt gamma-ray emissions and afterglow emissions (especially the early afterglow
emissions) may be a powerful way to test different jet models. Zhang et al. (2004), Lloyd-Roning et al. (2004), and Dai
\& Zhang (2005) showed that the observed $E_{\rm iso}-E_{\rm p}$ relation (Amati et al. 2002; Sakamoto et al. 2004;
Lamb et al. 2005; Liang, Dai \& Wu 2004), $\theta$ distribution, GRB distributions in ($\theta$, z)-plane and in
($E_{\rm p}$, $S_{\gamma}$)-plane, luminosity function, and $\log N- \log P$ distribution are roughly consistent with
simulation results based on a quasi-universal Gaussian-like jet model, where $E_{\rm iso}$ is the equivalent-isotropic
energy of GRBs, $S_{\gamma}$ is the observed fluence in gamma-ray band, and $z$ is the redshift of GRBs. Firmani et al.
(2004) constrained the isotropic luminosity function and formation rate of long GRBs by fitting models jointly to both
the observed differential peak flux and redshift distributions, and found evidence supporting a jet model intermediate
between universal power-law jet model and quasi-universal Gaussian structured model. For the uniform jet model their
result is compatible with an angle distribution between $2^{o}$ and $15^{o}$. The USJ model has a power to predict the
distribution of viewing angles, $P^{\rm th}(\theta)$. Perna, Sari, \& Frail (2003) utilized this power to test the USJ
models, and found that $P^{\rm th}(\theta)$ is roughly consistent with the observed one derived from a sample of 16
events with $\theta$ known from Bloom et al. (2003). However, Nakar, Granot \& Guetta (2004) made a further analysis,
and found that $P^{\rm th}(\theta, \ z)$ does not agree with the observed result. Liang, Wu \& Dai (2004) tested the
USJ model by simulations based on the $E_{\rm{iso}}-E_{\rm p}$ relation  and the assumption of a standard energy budget
in GRB jets (Frail et al. 2001). In their simulations, they adopted a high threshold of gamma-ray fluence, and found
that simulated $P(\theta)$ and $P(\theta, \ z)$ are consistent with USJ model predictions.

The largest GRB sample available so far is the BATSE GRB sample. Statistical test with this sample might be more
reasonable and reliable. More recently, Ghirlanda, Ghisellini, \& Lazzati (2004) discovered a very tight correlation
between the gamma-ray energy in GRB jet ($E_{\gamma}$) and $E_{\rm p}^{'}$, where $E_{\rm p}^{'}=E_{\rm p}\times(1+z)$
(hereafter GGL-relation). It gives an empirical way to estimate the $\theta$ of a burst once the $E_{\rm p}^{'}$ of the
bursts is available. In this analysis, we use this relationship to derive the empirical $\theta$ distribution of the
long BATSE GRB sample, $P^{\rm em}(\theta)$, and the distribution of these GRBs in the ($\theta, \ z$)-plane, $P^{\rm
em}(\theta, \ z$). We test different USJ models by comparing the model predictions, $P^{\rm th}(\theta)$ and $P^{\rm
th}(\theta, \ z)$, with our empirical results. The differences of this analysis from both Perna et al. (2003) and Nakar
et al. (2004) are summarized as follows. (1) The sample we used is the long BATSE GRB ($T_{90}>2$ seconds) sample (1213
bursts) but not the current $\theta$-known GRB sample. The current $\theta$-known GRB sample is too small and it
suffers greatly observational biases and sample selection effects (e.g., Broom \& Leob 2002; Bloom 2003). (2) The
observed GRB rate is taken as a corrected star formation rate proposed by Bloom (2003). When we compute the $\theta$
for each burst, we assign a redshift from the observed GRB rate model by a simple simulation. This GRB rate model is
also used to calculate the $P(\theta)$ and $P(\theta, \ z)$ predicted by the USJ models. (3) The $\theta$ of each burst
is calculated by the GGL-relation. Since the $E_{\rm p}$ values of most of the bursts in our sample are not available,
we derive the $E_{\rm p}$ values from a relationship between $E_{\rm p}$ and hardness ratios ($HR$). (4) To ensure that
the sample should be regarded as a complete one and to make the comparisons between empirical results and model
predictions reasonable, we use an ``effective" threshold according to our sample selection criteria. (5)Various USJ
models have been tested.

This paper is arranged as follows. In section 2, we describe our empirical approach to derive $P^{\rm em}(\theta)$ and
$P^{\rm em}(\theta, z)$ for the long BATSE GRBs. In section 3, we present the theoretical models of $P^{\rm
th}(\theta)$ and $P^{\rm th}(\theta, z)$ predicted by the USJ models. Results are presented in section 4, and
discussion and conclusions are presented in section 5. Throughout this paper we adopt $\Omega_{\rm M}=0.3$,
$\Omega_\Lambda=0.7$, and $H_0=71$ km Mpc$^{-1}$ s$^{-1}$.

\section{Empirical Approach}
The energy release in the gamma-ray band of a GRB jet is
\begin{equation}\label{ejet}
E_{\gamma}=E_{\rm iso}(1-\cos\theta),
\end{equation}
and $E_{\rm iso}$ is calculated by
\begin{equation}\label{eiso}
E_{\rm iso}=\frac{4 \pi D _L ^2 kS_{\gamma}}{1+z},
\end{equation}
where $S_{\gamma}$ is the observed gamma-ray fluence, $D_L$ is the luminosity distance at redshift $z$, and $k$ is a
factor to correct the observed fluence in an instrument band to a standard bandpass in the rest-frame ($1-10^4$ keV in
this analysis; Bloom et al. 2001). The GGL-relation is,
\begin{equation}\label{ej50}
E_{\gamma,50}\simeq [E_{\rm{p},2}(1+z)]^{1.5},
\end{equation}
where $E_{p,2}=E_{p}/100$ keV and $E_{\gamma, 50}=E_{\gamma}/10^{50}$ ergs (see also, Dai, Liang, \& Xu 2004). Although
the physics behind this relationship is still poorly understood, the very small dispersion of this relationship makes
it a reliable way to estimate the value of $\theta$, once the $E_{\rm p}$ and $z$ of a burst are well measured. We use
this relationship to calculate the $\theta$ values of the long BATSE GRBs. Combining Eqs. \ref{ejet}-\ref{ej50}, we
have
\begin{equation}\label{theta}
\theta=\arccos[1-\frac{1}{4\pi}\frac{E_{\rm{p},2}^{1.5}(1+z)^{2.5}}{D_{L,28}^{2}S_{\gamma,-6}k}].
\end{equation}

The $E_{p}$ values of the bursts in our sample are only available for some bright GRBs (Band et al. 1993; Ford et al.
1995; Preece et al. 1998, 2000). It is well known that the distribution of $E_{\rm p}$ and $HR$ of long BATSE GRBs are
narrowly clustered (e.g., Preece et al. 2000). In a previous work we showed that the $E_{\rm p}$ and $HR$ of those GRBs
with a moderate $E_p$ ($100\sim 1000$ keV) and $HR$ ($1.6\sim 6$, calculated by the fluence measured in the energy band
110-300 keV to that in 55-110 keV) are strong correlated (Cui, Liang, \& Lu 2005). Thus, we estimate the $E_{\rm p}$
values by the correlation between $E_{\rm p}$ and $HR$. We show this correlation for a bright GRB sample of 149 bursts
presented by Lloyd-Ronning \& Ramirez-Ruiz (2002) in the upper panel of Figure 1. We perform a linear least square fit
at $1\sigma$ confidence level to the two quantities. We have
\begin{equation}\label{ephr}
\log E_{\rm{p}}=(1.86\pm 0.14)+(1.16\pm 0.09) \log HR
\end{equation}
with a linear correlation coefficient of 0.74 and a chance probability of $p< 10^{-4}$. The reduced $\chi^2$ is 3.84.
We mark the $1\sigma$ region in Figure 1 (gray band). The distribution of the $HR$ for the long BATSE GRB sample is
also shown in the lower panel of Figure 1. From Figure 1 we find that most of the GRBs in our sample well follow this
relationship.

The redshifts of most GRBs in our sample are unknown. Please note that our purpose is to examine whether or not $P^{\rm
em}(\theta)$ and $P^{\rm em}(\theta,z)$ are consistent with those predicted by the USJ models. A GRB rate model has to
be used in our calculations. To compare the model predictions with the empirical results, the same GRB rate model
should be used to derive the empirical results and model results. We assume that the observed GRB rate as a function of
redshift for the long GRBs is the same as that derived from the currently redshift-known GRB sample. Bloom (2003)
considered a correction factor of $D_L^{-2}$ for $z>1$ and showed that the models of star formation rate as a function
of redshift, SF1, SF2, and SF3, from Porciani \& Madau (2001), are consistent with the observed GRB rate. We hence use
the same GRB rate model suggested by Bloom (2003) in our calculations. Since the largest redshift of GRBs observed so
far is 4.5, we also limit $z\leq 4.5$. For a given burst, its redshift is assigned by a simple Monte Carlo simulation.
The procedure of our simulation is as follows: (1) obtain the differential probability of the observed redshift
distribution with a bin size of 0.01, $dQ(z)/dz$; (2) derive the accumulated probability distribution of redshift,
$Q(z)$ ($0<Q(z)\leq 1$); (3) generate a random number, $m$ ($0<m\leq 1$); and (4) assign a $z$ value to the bursts by
$z=(z_{i+1}+z_{i})/2$, where $Q(z_i)<m$ and $Q(z_{i+1})>m$. Please note that the redshift assigned by this way for a
burst can be used only for a statistical purpose. Such a simulation ensures that $P(\theta)$ and $P(\theta,z)$
predicted by the USJ models and derived from BASTE observations are based on the same GRB rate model.

Based on the analysis above, we calculate the $\theta$ values by Eq. \ref{theta}. The uncertainty of $\theta$ is

\begin{equation}\label{}
\sigma_{\theta}=\sqrt{(\frac{\partial \theta}{\partial E_{\rm p}})^2\sigma_{E_{\rm p}}^2+(\frac{\partial
\theta}{\partial E_{\rm iso}})^2\sigma_{E_{\rm iso}}^2} =\frac{y}{\sqrt{1-x^2}}\sqrt{(1.5\frac{\sigma_{E_{\rm
p}}}{E_{\rm p}})^2+(\frac{\sigma_{E_{\rm iso}}}{E_{\rm iso}})^2},
\end{equation}
where $y=[E_{p,2}(1+z)]^{1.5}/E_{iso,50}$, $x=1-y$. The uncertainties of $E_{\rm iso}$ and $E_{\rm p}$ are given by
\begin{equation}
\sigma_{E_{\rm iso}}=\frac{4 \pi D _L ^2}{1+z}k\sigma_{S_{\gamma}}
\end{equation}
and
\begin{equation}\label{}
\sigma_{E_{\rm p}}=\sqrt{(\frac{\partial E_{\rm p}}{\partial a })^2\sigma_a^2+(\frac{\partial E_{\rm p}}{\partial
b})^2\sigma_b^2+(\frac{\partial E_{\rm p}}{\partial HR})^2\sigma_{HR}^2}=E_{\rm p}\ln10\sqrt{\sigma_a^2+(\log
HR)^2\sigma_b^2+(\frac{b\sigma_{HR}}{HR\ln 10})^2},
\end{equation}
respectively, where $a=1.86\pm 0.14$ and $b=1.16\pm 0.09$. Please note when we calculate the $\sigma_{E_{\rm iso}}$,
the uncertainty of $k$ is ignored since the spectral parameters of the bursts are not available (we assume $\alpha=-1$
and $\beta=-2.3$ to compute $k$ values for all the bursts).

\section{Theoretical Models}

The $P(\theta)$ and $P(\theta,z)$ predicted by the USJ  models for a given detection threshold are
\begin{equation}
P^{\rm th}(\theta)=\sin \theta\int^{z_{max}}_0dz\frac{R_{\rm{GRB}}}{1+z}\frac{dV}{dz}
\end{equation}
and
\begin{equation}
P^{\rm th}(\theta,z)=\sin\theta \frac{R_{GRB}}{1+z}\frac{dV}{dz},
\end{equation}
respectively, where $R_{GRB}$ is the GRB rate per unit comoving time per unit comoving volume, $dV/dz$ is the comoving
volume element at $z$, and $z_{max}$ is the maximum redshift up to which a burst with viewing angle $\theta$ can
satisfy the detection threshold. In this analysis, the $R_{GRB}$ is taken as the observed GRB rate suggested by Bloom
(2003). The $z_{max}$ is determined by instrument sensitivity and jet structure model. It can be derived from
\begin{equation}
E_{\gamma}=\int^{\pi/2}_{0}4\pi \epsilon (\theta)\sin\theta d\theta=\frac{{4\pi
D_{L}^2(z_{max})\tilde{S}_{th}}k}{1+z_{max}}(1-\cos\theta),
\end{equation}
where $\tilde{S}_{th}$ is the ``effective" threshold of the long GRB sample observed by BATSE, and $\epsilon(\theta)$
is energy density in per solid angle as a function of $\theta$. Please note that $\tilde{S}_{th}$ does not correspond
to the BATSE threshold. An observed GRB sample is significantly affected by observational biases and sample selection
effects, especially when the completeness at low fluxes is considered. The ``effective" threshold should correspond to
a given sample selection criterion to ensure that the sample should be regarded as a complete one in this threshold.
Such a sample should be selected from a sensitive all-sky survey. The fluence distribution of our sample is shown in
Figure 2. It is found 95\% of the bursts satisfy $S>3.2\ \times 10^{-7}$ ergs. cm$^{-2}$. We thus take $\tilde{S}_{th}$
as $3.2 \times 10^{-7}$ ergs. cm$^{-2}$. The energy density profile $\epsilon(\theta)$ for a single power-law jet,
Gaussian jet, and two-components jet are written as follows.

Power-law jet:

\begin{equation}
\epsilon(\theta)=\epsilon_c (\frac{\theta}{\theta_c})^{-2}
\end{equation}
where $\epsilon_c$ is the core energy density when $\theta<\theta_c$.

Gaussian jet:
\begin{equation}
\epsilon(\theta)=\epsilon_0 e ^{-\theta^2/2\theta_0^2}
\end{equation}
where $\theta_0$ is a characteristic width of the jet, and $\epsilon_0$ is the maximum value of the energy density.

Two-component jet:

\begin{equation}
\epsilon(\theta)=\epsilon_0 (e ^{-\theta^2/2\theta_{c,1}^2}+\lambda e ^{-\theta/2\theta_{c,2}^2})
\end{equation}
where $\theta_{c,1}$ and $\theta_{c,2}$ are respectively characteristic widths of the two components, and $\lambda$ is
the ratio of energy densities in the two components.

\section{Results}
Based on empirical approach and theoretical models discussed above, we calculate the empirical and theoretical $\theta$
distributions and the distributions of these GRBs in the ($\theta, \ z$)-plane. By comparing these empirical results
with model's predictions, we test these jet models.

Zhang et al. (2004) showed that the current GRB/XRF prompt emission/afterglow data can be described by a quasi-Gaussian
type  structured jet (or similar structure jet) with a typical opening angle of 0.1 rad and with a standard jet energy
of $\sim 10^{51}$ ergs. In addition, Liang (2004) showed that the fluence of those GRBs with $\theta<0.1$ rad is almost
a constant. We thus take $\theta_0=0.1$ rad and $\epsilon_0= 10^{51}$ ergs for the Gaussian jet. For the power-law jet
model, we take $\theta_c=0.1$ rad and let $\epsilon_c$ be variable. We find that $\epsilon_c=2.3\times 10^{50}$ ergs
yields the best consistency between $P^{\rm em}(\theta)$ and $P^{\rm th}(\theta)$. Berger et al. (2003) argued that the
observations of GRB 030329 require a two-component explosion with $\theta_{c,1}=0.1$ and $\theta_{c,2}=0.3$. Liang \&
Dai (2004) obtained the similar results based on the $E_{\rm p}$ distribution observed by BATSE and HETE-2. We thus
take $\theta_{c,1}=0.1$ and $\theta_{c,2}=0.3$ for the two-component model. The ratio of energy densities in the two
components, $\lambda$, is suggested to be $10^{-1.7}$ in Liang \& Dai (2004). In this analysis we let it be variable in
the range of (0.01, 0.1) to derive a good consistency between $P^{\rm em}(\theta)$ and $P^{\rm th}(\theta)$. We use
different values of $\lambda$, and find that $\lambda\sim 0.08$ gives the best consistency. The $\epsilon_0$ in the
two-component jet model is taken as $(10\pm 4)\times 10^{50}$ ergs, where the error is assumed to be normal
distributed. These model parameters are summarized in Table 1.

The empirical results and their comparisons to model's predictions are shown in Figs. (3)-(5). The comparisons between
$P^{\rm em}(\theta)$ and $P^{\rm th}(\theta)$ are shown in the left panels of the three figures, and comparisons
between $P^{\rm em}(\theta, \ z)$ and $P^{\rm em}(\theta, \ z)$ are shown in the right panels. The error bars of
$P^{\rm em}(\theta)$ are calculated by the following method: we first derive $P^{\rm em}(\theta+\sigma_\theta)$ and
$P^{\rm em}(\theta-\sigma_\theta)$, and then obtain $\sigma _{P^{\rm em}(\theta)}$ by $\sigma_{ P^{\rm
em}{(\theta)}}=|P^{\rm em}{(\theta)}(\theta+\sigma_\theta)-P^{\rm em}{(\theta)}(\theta-\sigma_\theta)|/2$. Those GRBs
with $\theta$-known from Bloom et al. (2003) are also marked by star symbols in the $(\theta, \ z)$-planes.  The
distribution of these GRBs are roughly consistent with the 2$\sigma$ region of the $P^{\rm em}(\theta, \ z)$. We
quantify the difference between $P^{\rm em}(\theta)$ and $P^{\rm th}(\theta)$ by a Kolmogorov-Smirnov (K-S) test. The
result of the K-S test is depicted by a statistic of $P_{K-S}$: a small value of $P_{K-S}$ indicates a significant
difference between the two distributions ($P_{K-S}=1$ means two distributions are identical, and $P_{K-S}<1.0\times
10^{-4}$ suggests that they are significantly different). The K-S test results are also listed in Table 1. For the
comparison between $P^{\rm em}(\theta, \ z)$ and $P^{\rm em}(\theta, \ z)$, we do not have a quantifiable way to
estimate the difference but evaluate it by eye instead.

From Fig. (3) and Table 1, we find the empirical $P^{\rm em}(\theta)$ is quite consistent with $P^{\rm th}(\theta)$
predicted by the power-law jet model. The K-S test shows $P_{K-S}=0.575$, strongly indicating an agreement between
$P^{\rm em}(\theta)$ and $P^{\rm th}(\theta)$. The $P^{\rm em}(\theta)$ distribution is well fitted by a log-normal
function,
\begin{equation}
P^{\rm em}(\theta)=\frac{0.67}{0.57\sqrt{\pi/2}}e^{-2(\frac{\log \theta+0.76}{0.57})^2}.
\end{equation}
However, the $P^{\rm em}(\theta, \ z)$ is greatly different from $P^{\rm th}(\theta, \ z)$ predicted by this jet model.
Even in $1\sigma$ regions, they are quite different. The lack of high-$z$ and large-$\theta$ (the right-top of the
right panel of Fig. 3) make most of this difference.  We adjust the parameters of this jet model, but we still do not
get an agreement between $P^{\rm em}(\theta, \ z)$ and $P^{\rm th}(\theta, \ z)$.

Shown in Fig. (4) is the comparisons of the empirical results with whose predicted by the Gaussian jet model. It is
found that the $P^{\rm th}(\theta)$ rapidly drops at $\theta>0.3$ rad. This is significantly different from the
empirical result. The K-S test for the two distributions shows $P_{K-S}=2.1\times 10^{-9}$, indicating that the
hypothesis of consistency is rejected. Similar to the $P^{\rm th}(\theta, \ z)$ predicted by this jet model, the
$P^{\rm th}(\theta, \ z)$ predicted by the Gaussian jet model is also deviates the $P^{\rm em}(\theta, \ z)$. However,
comparing the results shown in the right panels of Figures 3 and 4, one can find that the $P^{\rm th}(\theta, \ z)$
predicted by the Gaussian jet model is more consistent with $P^{\rm em}(\theta, \ z)$ than that predicted by the
power-law jet model.

The comparisons of $P^{\rm em}(\theta)$ and $P^{\rm em}(\theta, \ z)$ with those derived from the two-component jet
model are shown in Figure 5. The K-S test for the distributions of $P^{\rm em}(\theta)$ and $P^{\rm th}(\theta)$
obtains $3.55\times 10^{-4}$, indicating the consistence between the two distributions is marginally accepted. The
$P^{\rm th}(\theta, \ z)$ predicted by this jet model is also roughly consistent with $P^{\rm em}(\theta, \ z)$ in
1$\sigma$ region.

\begin{table}
\centering

\caption{The jet model parameters and the results of K-S test for $P^{\rm em}(\theta)$ and $P^{\rm th}(\theta)$}
\begin{tabular}{lccc}
\hline\hline
 Model&$\epsilon_c$ or $\epsilon_0$ ($10^{50}$erg)&$\theta_c$ or $\theta_0$ (rad)&$P_{K-S}$\nl
 \hline

Power-law&2.3 &0.1&0.575\nl

Gaussian &10&0.1&$2.10\times 10^{-9}$\nl

Two-component&$10\pm4$&0.1, 0.3&$3.55\times 10^{-4}$\nl

\hline
\end{tabular}
\end{table}

\section{Conclusions and Discussion}
Based on the assumption that the observed GRB rate is proportional to a corrected star formation rate, we have derived
the empirical distributions of long BATSE GRBs, $P^{\rm em}(\theta)$ and $P^{\rm em}(\theta, \ z)$, by the
GGL-relation. Our results show that $P^{\rm em}(\theta)$ is well fitted by a log-normal distribution centering at $\log
\theta=-0.76$ with a width of $0.57$. We test different USJ models by comparing model predictions, $P^{\rm th}(\theta)$
and $P^{\rm th}(\theta, \ z)$, with the empirical results. To make the comparison reasonable, an ``effective"
threshold, which corresponds to the sample selection criteria of the long GRB sample, is used. We find that $P^{\rm
th}(\theta)$ predicted by the power-law jet model is well consistent with $P^{\rm em}(\theta)$, but $P^{\rm th}(\theta,
\ z)$ predicted by this model is significantly different from $P^{\rm em}(\theta, \ z)$. Inversely, the $P^{\rm
th}(\theta, \ z)$ predicted by a single Gaussian jet model is more consistent with $P^{\rm em}(\theta, \ z)$ than that
predicted by the power-law jet model, but the $P^{\rm th}(\theta)$ predicted by this model rapidly drops at
$\theta>0.3$ rad, which greatly deviates $P^{\rm em}(\theta)$. Both $P^{\rm th}(\theta)$ and $P^{\rm th}(\theta, \ z)$
predicted by a jet model with two Gaussian components are roughly consistent with $P^{\rm em}(\theta)$ and $P^{\rm
em}(\theta, \ z)$.

The structure of jet is crucial to understanding the nature of GRBs, such as their rate, luminosity function, and
explosion energy. Numerical simulations of jet propagation within collapsar context by Zhang, Woosley, \& Heger (2004b)
showed that the jet structure seems to be quite complicated. A universal, single-component jet seems to be difficult to
describe such a jet structure, and a jet with two (or more) components might be more reasonable. The observations of
GRB 030329 (Price et al. 2003; Berger et al. 2003; Sheth et al. 2003) and the bimodal distribution of the observed
$E_{\rm p}$ of HETE-2 bursts (Liang \& Dai 2004) are evidence for this jet structure. The results of this work also
seem to support this jet model. Zhang et al. (2004a), Lloyd-Ronning et al. (2004), and Dai \& Zhang (2005) showed that
quasi-universal Gaussian jet can well interpret the observations of the current well follow-up GRB sample. Please note
that this sample is a bright GRB sample. The median of the fluence in this sample is $\sim 2.5\times 10^{-5}$ erg
cm$^{-2}$ (Bloom et al. 2003). It is possible that the emissions of these bursts are dominated by the core component of
the jet. Thus, a quasi-universal Gaussian jet might well explain the observations of these bursts.

The true GRB rate as a function of redshift is difficult to establish from the current GRB sample. It is generally
suggested that the GRB rate follows the star formation rate. Various models of the star formation rate are presented in
the literature. Whether or not the model of the star formation rate affects our results?  We use a model of the
observed GRB rate suggested by Bloom (2003), who constructed a correction factor of $D_L^{-2}$ when $z>1$. This factor
leads to the observed GRB rate deeply decay when $z>1$. Thus, different models of the star formation rate may give
almost the same observed GRB rate (Bloom 2003), and then the model of the star formation rate does not significantly
affect our results. In fact, we focus on the comparison of both results on theoretical bases and on the observational
bases. The results of the comparison should not be greatly affected by the model of star formation rate (Liang, Wu \&
Dai 2004).

Our empirical results are derived by the GGL-relation. This relationship depends on cosmological parameters. In this
work we adopt $\Omega_M=0.3$ and $\Omega_{\Lambda}=0.7$. We check if the cosmological parameters significantly affect
the results of the comparisons between our empirical results and model predictions. We take $\Omega_M=0.5$ and
$\Omega_{\Lambda}=0.5$. In this case, the GGL-relation becomes $E_{\gamma,50}=(0.90\pm0.12) E_{\rm p}^{1.42\pm 0.10}$
for a GRB sample presented by Xu et al. (2005). We show the comparison between the $P^{\rm em}(\theta)$ based on this
relationship and $P^{\rm th}(\theta)$ predicted by the power-law jet model in the cosmology model with $\Omega_M=0.5$
and $\Omega_{\Lambda}=0.5$ in Figure 6. The K-S test for the two distributions shows $P_{K-S}=0.123$, confidently
suggesting that they are consistent. Comparing the results shown in Figure 6 to that shown in the left panel of Figure
3, one can find that cosmological parameters do not significantly affect our results.

We should clarify that our empirical approach and theoretical model are independent without toutology. The model
predictions are statistical distributions, while the empirical results are based on the relationships related to the
spectral properties and energy release of GRBs. The empirical approach and theoretical model are intrinsically
different.

Our empirical results are roughly consistent with the results from currently $\theta$-known GRB sample. This is a
self-consistent result because the GGL-relation was discovered by this GRB sample. For the bursts in this sample their
peak energies, temporal breaks of optical afterglow light curves, and redshifts are well measured. Such a sample must
suffers greatly observational biases and sample selection effects, especially when the completeness at low fluxes and
the bias of redshift measurement are considered. Band \& Preece (2005) argued that the GGL-relation may be  an artifact
of the selection effects, and these selection effects may favor sub-populations of GRBs. If it is really the case, the
selection effects should affect our empirical results.

We thank the anonymous referee for his/her valuable suggestions and comments. We also thank Bing Zhang, Zigao Dai, and
Yiping Qin for their helpful discussions. This work was supported by the National Natural Science Foundation of China
(Grants 10463001).

\begin{figure}
\plotone{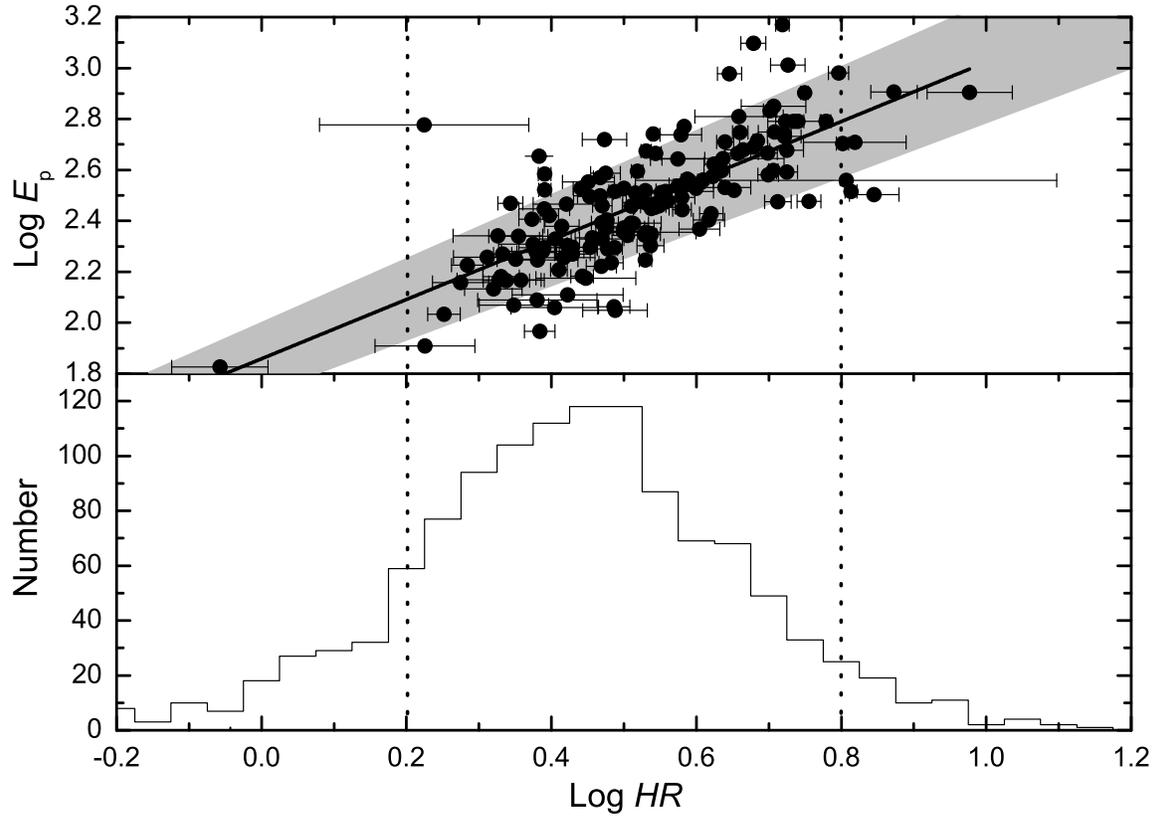}\caption{$Upper$: log $E_{\rm p}$ as a function of  $\log HR$ for a sample of 149 bright GRBs
presented by Llyod-Ronning \& Ramirez-Ruiz (2002). The solid line is the best fit line, and the grey band marks the
1$\sigma$ region of the best fit. $Lower$: The number distribution of $\log HR$ for 1213 long BATSE GRBs. Two
vertical-dotted lines mark a region within which bursts follow the $E_{\rm p}-HR$ relationship shown in the upper
panel. \label{fig1}}
\end{figure}

\begin{figure}
\plotone{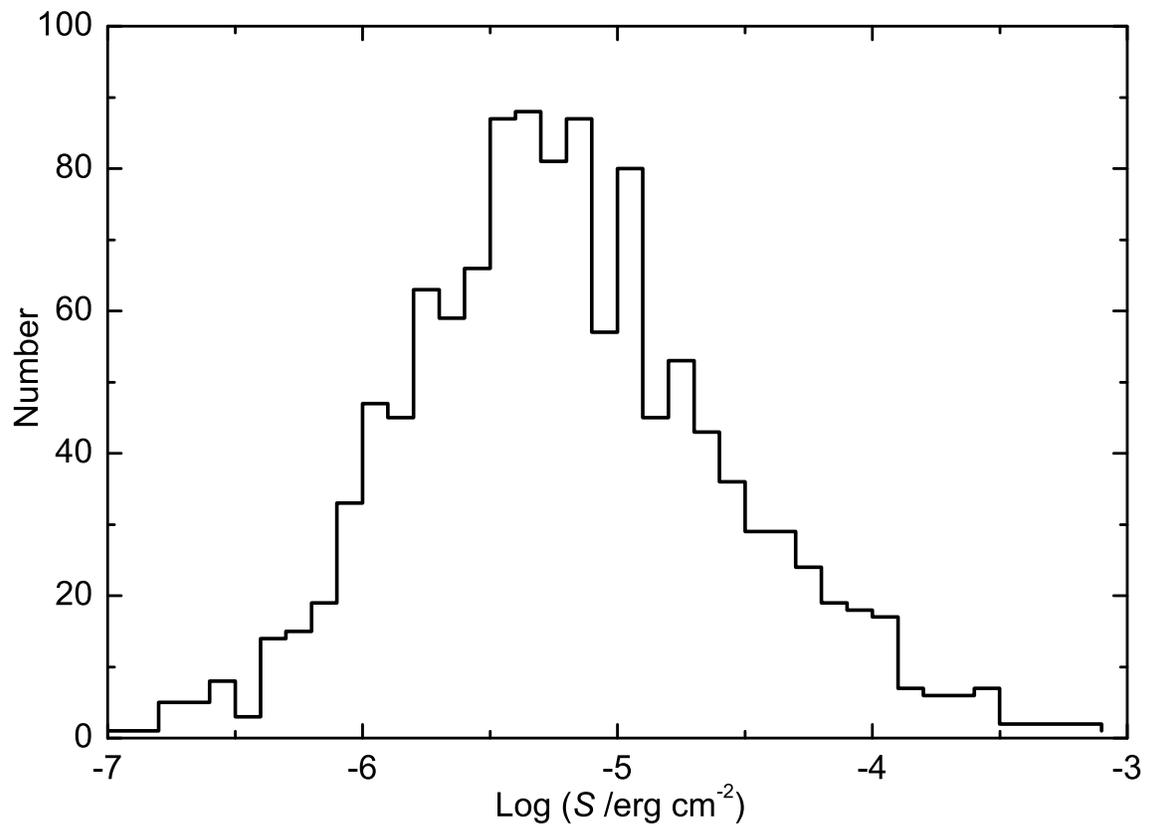} \caption{Number distribution of the observed fluence in energy band $25-2000$ keV for 1213 long
BATSE GRBs. \label{fig2}}
\end{figure}

\begin{figure}
\plotone{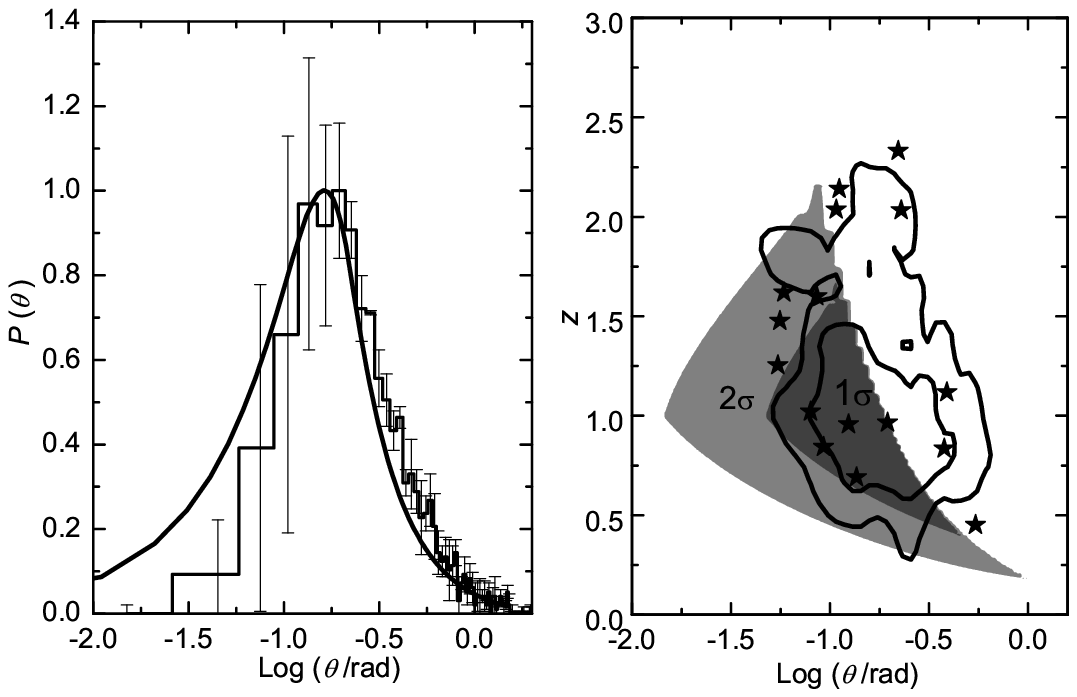} \caption{Comparisons between the GRB probability distributions, $P(\theta)$ (left panel) and
$P(\theta, \ z)$ (right panel), obtained by our empirical approach and by the power-law jet model. In the $left$ panel,
the step line with error bars is our empirical results, and the solid line is the result of the model. The line
contours in the $right$ panel are our empirical $P(\theta, \ z)$ (1 $\sigma$ and 2 $\sigma$) and the grey contours are
the results of the jet model. The stars in the $right$ panel are the distribution of the bursts with $\theta$ being
derived by jet break times (from Bloom et al. 2003).\label{fig3}}
\end{figure}

\begin{figure}
\plotone{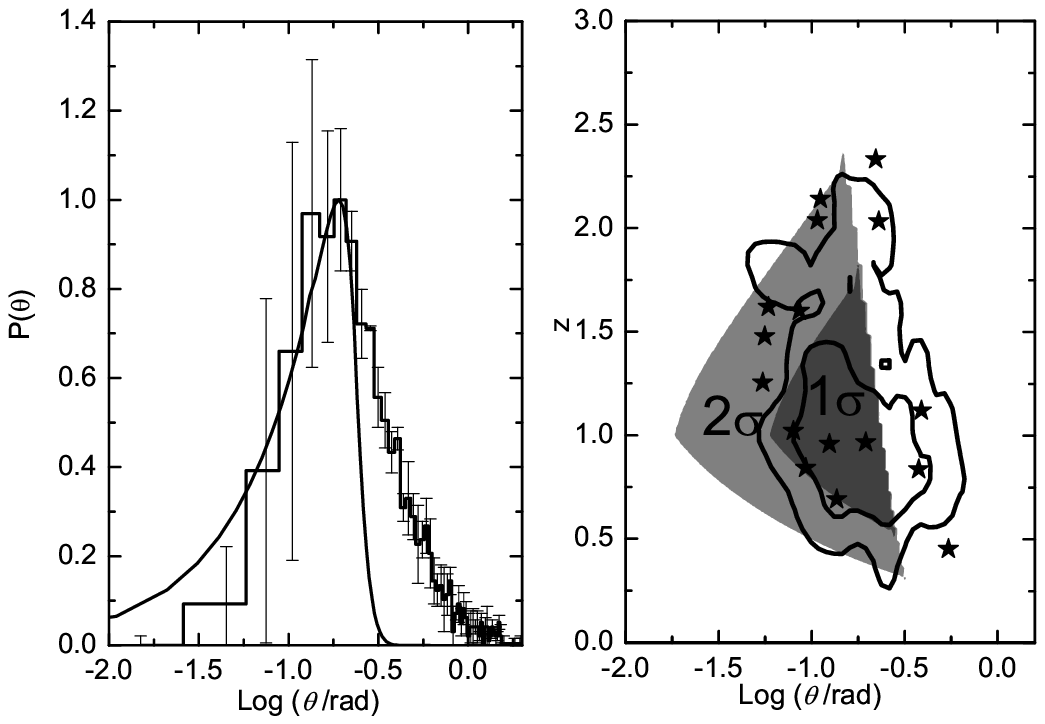} \caption{Comparisons between the GRB probability distributions, $P(\theta)$ (left panel) and
$P(\theta, \ z)$ (right panel), obtained by our empirical approach and by the Gaussian jet model. In the $left$ panel,
the step line with error bars is our empirical results, and the solid line is the result of the model. The line
contours in the $right$ panel are our empirical $P(\theta, \ z)$ (1 $\sigma$ and 2 $\sigma$) and the grey contours are
the results of the jet model. The stars in the $right$ panel are the distribution of the bursts with $\theta$ being
derived by jet break times (from Bloom et al. 2003). \label{fig4}}
\end{figure}

\begin{figure}
\plotone{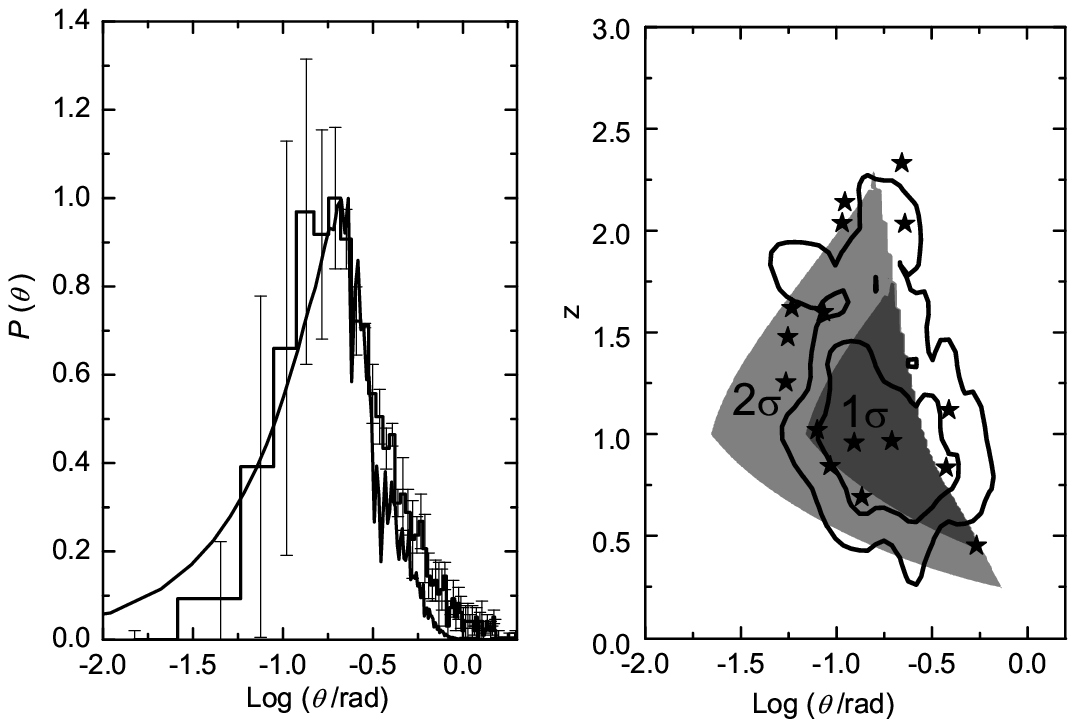} \caption{Comparisons between the GRB probability distributions, $P(\theta)$ (left panel) and
$P(\theta, \ z)$ (right panel), obtained by our empirical approach and by the two-component jet model. In the $left$
panel, the step line with error bars is our empirical results, and the solid line is the result of the model. The line
contours in the $right$ panel are our empirical $P(\theta, \ z)$ (1 $\sigma$ and 2 $\sigma$) and the grey contours are
the results of the jet model. The stars in the $right$ panel are the distribution of the bursts with $\theta$ being
derived by jet break times (from Bloom et al. 2003). \label{fig5}}
\end{figure}

\begin{figure}
\plotone{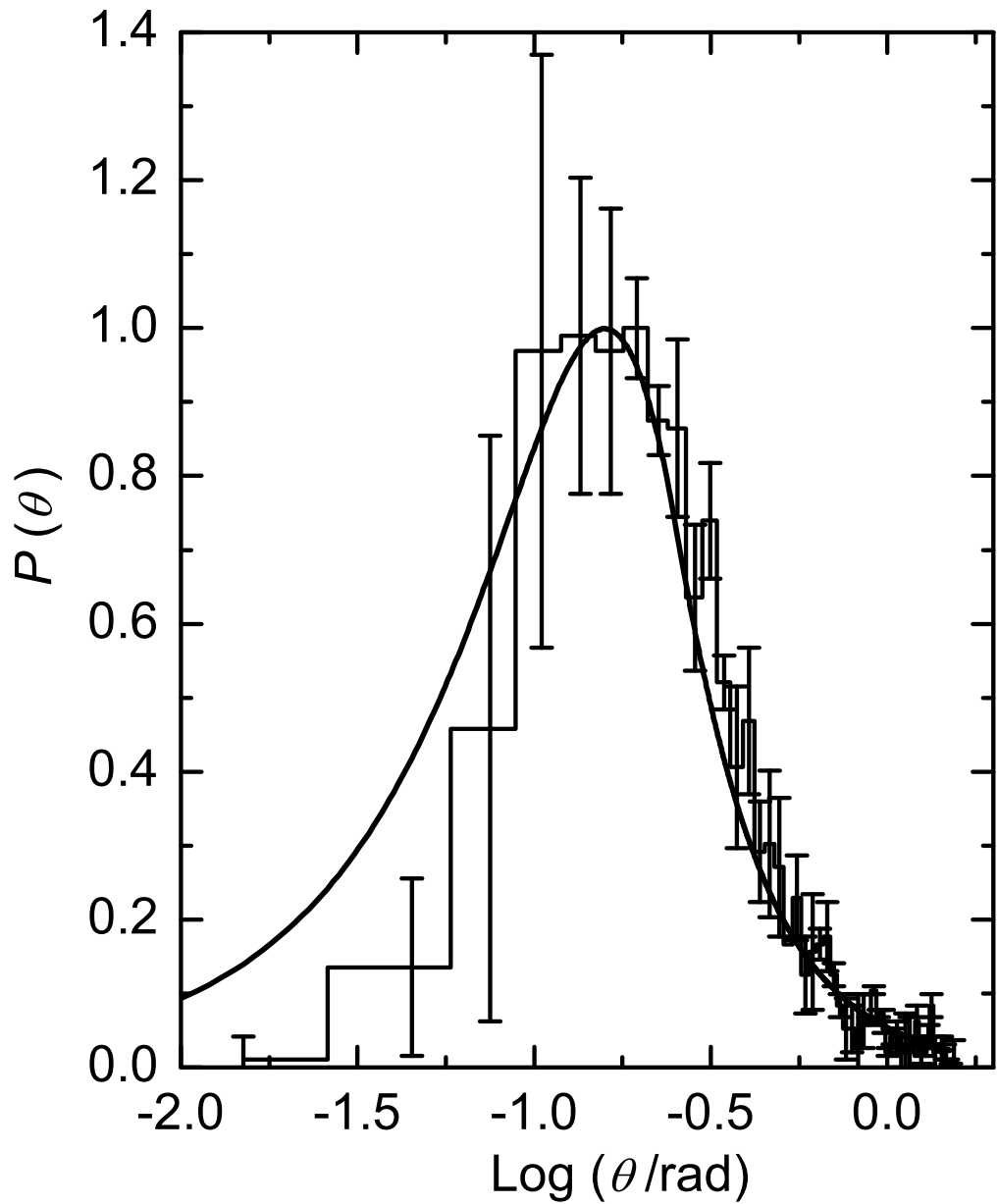} \caption{Comparisons between the $\theta$ distributions derived by our empirical approach (step line
with error bars) and by the power-law jet model (solid line) in a cosmology with parameters $\Omega_{\rm M}=0.5$ and
$\Omega_{\Lambda}=0.5$. \label{fig6}}
\end{figure}

\end{document}